\documentclass[12pt]{article}
\pdfoutput=1
\usepackage{jheppub}
\usepackage{amsfonts}
\usepackage{amscd}
\usepackage{amssymb}
\usepackage{amsmath,bbm}
\usepackage{epsfig}
\usepackage{latexsym}
\usepackage{mathtools}
\usepackage{arydshln}
\usepackage{tikz-cd}
\usepackage{hyperref}
\usepackage{color}
\usepackage{subfig}
\usepackage{slashed}
\usepackage{euscript, mathrsfs}
\usepackage[vcentermath]{youngtab}
\usepackage{marginnote}
\usepackage{bm}

\def\be{\begin{equation}}
\def\ee{\end{equation}}
\def\bea{\begin{eqnarray}}
\def\eea{\end{eqnarray}}
\def \beaa {\begin{equation}\begin{aligned}}
\def \eeaa {\end{aligned}\end{equation}}

\newcommand{\nn}{\nonumber}

\newcommand\diff{\mathrm{d}}

\newcommand{\ii}{\mathrm{i}}

\newlength{\sswidth}

\def \be  {\begin{equation}}
\def \ee  {\end{equation}}
\def \ba  {\begin{eqnarray}}
\def \ea  {\end{eqnarray}}
\def \bb  {\begin{thebibliography}}
\def \eb  {\end{thebibliography}}
\def \lab #1 {\label{#1}}

\newcommand\cA{\mathcal{A}}

\newcommand\cK{\mathcal{K}}
\newcommand\cL{\mathcal{L}}
\newcommand\cM{\mathcal{M}}
\newcommand\cN{\mathcal{N}}

\newcommand\cR{\mathcal{R}}
\newcommand\cS{\mathcal{S}}
\newcommand\cT{\mathcal{T}}

\newcommand\cV{\mathcal{V}}

\newcommand\tr{\mathrm{Tr}}

\newcommand\ie{\textit{i.e.}}

\title{4d $\mathcal{N} = 1$/2d Yang-Mills Duality in Holography}
\author{Martin Fluder}
\affiliation{
Walter Burke Institute for Theoretical Physics,
California Institute of Technology\\
Pasadena, CA 91125, USA
}

\emailAdd{fluder@caltech.edu}

\preprint{CALT-TH-2017-057}

\abstract{
We study the supergravity dual of four-dimensional ${\mathcal{N}=1}$ superconformal field theories arising from wrapping M5-branes on a K\"ahler two-cycle inside a Calabi-Yau threefold. We derive an effective three-dimensional theory living on the cobordism between the infrared and ultraviolet Riemann surfaces, describing the renormalization group flows between AdS$_7$ and AdS$_{5}$ as well as between different AdS$_{5}$ fixed points. The realization of this system as an effective theory is convenient to make connections to known theories, and we show that upon imposing (physical) infrared boundary conditions, the effective three-dimensional theory further reduces to two-dimensional $SU(2)$ Yang-Mills theory on the Riemann surface.
}

\begin{document}
\maketitle
\bibliographystyle{JHEP}

\section{Introduction}

A fruitful perspective on a large class of four-dimensional superconformal field theories is to compactify the six-dimensional $\cN=(2,0)$ theory, which arises as the low energy effective worldvolume theory on a stack of M5-branes, on a Riemann surface~\cite{Gaiotto:2009we,Gaiotto:2009hg}. In order to preserve some supersymmetry, such theories are engineered by imposing a partial topological twist in six-dimensions. An interesting way to study such dimensional reductions is by considering them as renormalization group flows on the holographic dual supergravity side from the ultraviolet AdS$_{7}$ to the infrared AdS$_{5}$ geometry~\cite{Maldacena:2000mw}. In practice, one imposes the partial topological twist holographically in the ultraviolet AdS$_{7}$ regime (corresponding to the dual of the $\cN=(2,0)$ theory), thus allowing for arbitrary metric on the Riemann surface. Upon evolving along the renormalization group flow to the infrared AdS$_{5}$ fixed point one can either leave the topological twist manifest or relax that assumption in the bulk of the flow (\ie~only set it as a ultraviolet boundary condition). The latter approach was employed in~\cite{Anderson:2011cz} to prove that for particular types of flows, the metric on the Riemann surface ``smoothes out" to a constant curvature metric in the infrared.\footnote{See also~\cite{Fluder:2017nww}, where a similar ``holographic uniformization" was studied for M5-branes wrapping a particular class of K\"ahler four-cycles.}

In this paper, we study the particular setup of a stack of M5-branes wrapping a genus-$g$ Riemann surface $\Sigma_{g}$, giving rise to $\cN=1$ superconformal field theories in four dimensions. On the field theory side, this corresponds to a reduction of the six-dimensional $\cN=(2,0)$ theory on $\Sigma_{g}$ with a partial topological twist preserving $\cN=1$ supersymmetry. The corresponding M-theory setup is given by the M5-branes wrapping a calibrated K\"ahler two-cycle inside a Calabi-Yau threefold. Locally, the corresponding Calabi-Yau threefold can be described as the total space of a complex rank-two vector bundle $\cV_{\mathbb{C}}$ over the Riemann surface 
\be
\begin{tikzcd}
  \mathbb{C}^{2} \arrow[r] 
    & \cV_{\mathbb{C}} \arrow[d] \\
&\Sigma_{g} \end{tikzcd}
\ee
with $U(2)$ structure group. In the holographic supergravity approximation, this system was studied in~\cite{Bah:2011vv,Bah:2012dg}, by reducing the bundle $\cV_{\mathbb{C}}$ to two complex line bundles $\cL_{1}$ and $\cL_{2}$ over the Riemann surface $\cL_{1} \oplus \cL_{2} \to \Sigma_{g}$. This breaks the $U(2)$ structure group down to its maximal torus. In supergravity, this results in truncating the gauged supergravity theory down to the Abelian $U(1)\times U(1)$ theory of~\cite{1999PhLB..457...39L}, and in~\cite{Bah:2011vv,Bah:2012dg}, they work out the corresponding renormalization group flows from the ultraviolet AdS$_{7}$ (with slices at constant radius given by $\mathbb{R}^{3,1} \times \Sigma_{g}$) to the infrared AdS$_{5}\times \Sigma_{g}$ fixed points. Furthermore, they study and compare various quantities from the supergravity and field theory perspective. Subsequently, it was found in~\cite{Beem:2012yn}, that those quantities match objects in two-dimensional $SU(2)$ Yang-Mills theory on the same Riemann surface and its Morse theory treatment~\cite{Atiyah:1982fa}, where the group $SU(2)$ was argued to be precisely the structure group of $\cV_{\mathbb{C}}$, which is reduced to $SU(2)$ upon imposing that the total space is Calabi-Yau.\footnote{There is no obvious relation between this statement and the correspondence between the four-dimensional $\cN=2$ superconformal index and two-dimensional $q$-deformed Yang-Mills~\cite{Gadde:2011ik}.} One goal of the current paper is to solidify and generalize this connection, by explicitly deriving the $SU(2)$ two-dimensional Yang-Mills theory at the infrared AdS$_{5}$ fixed point from a very general supergravity setup.

\begin{figure}[h!]
\centering
\includegraphics[width=.85\textwidth]{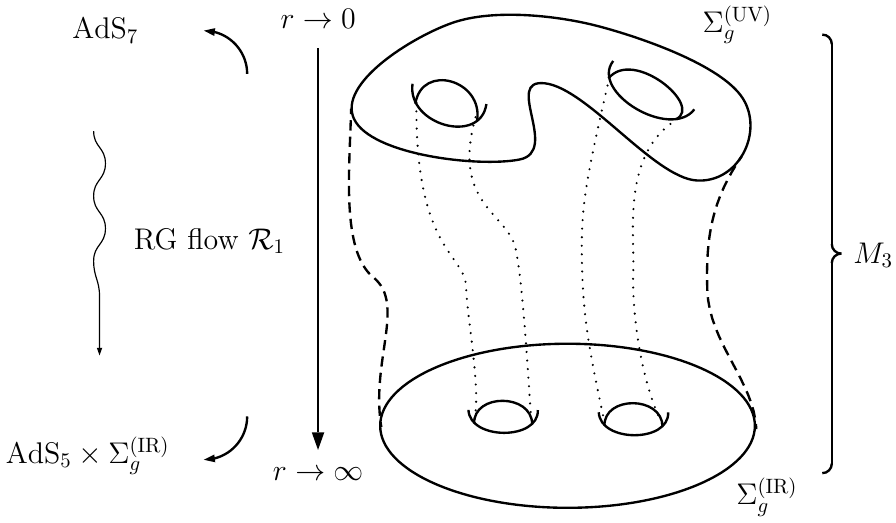}
\caption{ A schematic sketch of the holographic renormalization group flow $\cR_{1}$ from the ultraviolet (top) AdS$_{7}$, whose slices at fixed radial coordinate $r$ are given by $\mathbb{R}^{3,1}\times \Sigma_{g}^{(\rm UV)}$, to the infrared (bottom) AdS$_{5}\times \Sigma_{g}^{(\rm IR)}$ fixed points. The cobordism, $M_3$ is given by the evolution of the geometry of the Riemann surface along the flow. Furthermore, as showed in~\cite{Anderson:2011cz}, the metric on the Riemann surface ``uniformizes" in the infrared.}
\label{Fig:RGUVIR}
\end{figure}

\begin{figure}[h!]
\centering
\includegraphics[width=.85\textwidth]{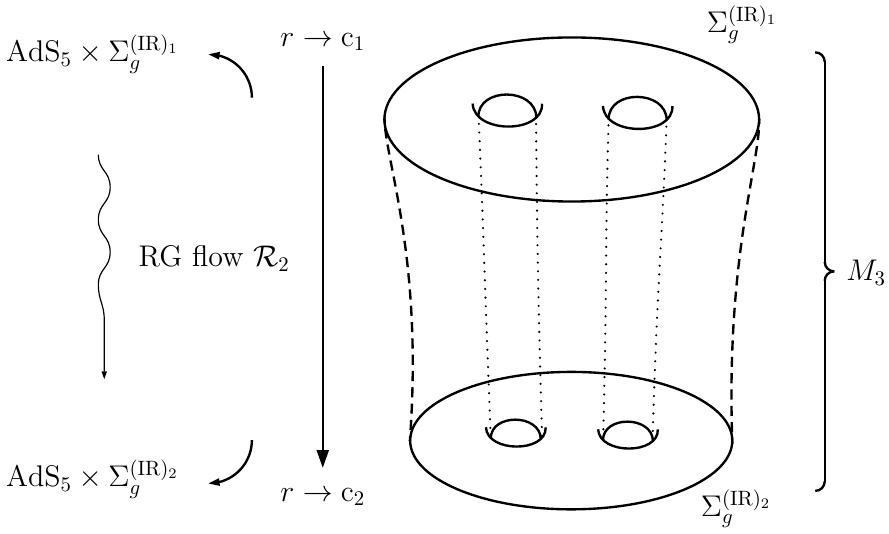}
\caption{A very schematic sketch of the holographic renormalization group flow $\cR_{2}$ between different AdS$_{5}\times \Sigma_{g}^{({\rm IR})_{i}}$ fixed points. This case relates two infrared theories and thus the Riemann surfaces $\Sigma_{g}^{({\rm IR})_{i}}$ are already uniform.}
\label{Fig:RGIRIR}
\end{figure}

For the purpose of this paper, we study this system of M5-branes wrapping a Riemann surface $\Sigma_{g}$ from an alternative angle, motivated to some extent by the AGT-correspondence~\cite{Alday:2009aq}. Namely, instead of treating it as an eleven-dimensional M-theory or seven dimensional effective gauged supergravity, we reduce the theory to an effective three-dimensional theory, which naturally appears upon imposing a physically well-motivated ansatz. The three dimensional theory lives on the cobordism $M_{3}$ given by the radial (renormalization group flow) direction $r$ times the Riemann surface (see Figure~\ref{Fig:RGUVIR} and Figure~\ref{Fig:RGIRIR}). This effective three-dimensional ``cobordism theory", for which we can write down an explicit Lagrangian, then describes the geometry and fields of the evolution of a general class of holographic renormalization group flows. For instance, it describes the renormalization group flow $\cR_{1}$ of Figure~\ref{Fig:RGUVIR}, between the ultraviolet (asymptotically twisted) AdS$_{7}$ solution and the infrared AdS$_{5}$ fixed points. In general, the metric on the Riemann surface in the ultraviolet $\Sigma^{(\rm UV)}_{g}$ can be picked arbitrarily, since one imposes a topological twist asymptotically, guaranteeing that we have proper supersymmetric solutions. In the infrared, the Riemann surface $\Sigma^{(\rm IR)}_{g}$ is then expected to be ``smoothed out" or ``uniformized" as compared with $\Sigma^{(\rm UV)}_{g}$~\cite{Anderson:2011cz}. Another example of an interesting renormalization group flow $\cR_{2}$, also described by the effective three-dimensional ``cobordism theory", is sketched in Figure~\ref{Fig:RGIRIR}, interpolating between different AdS$_{5}$ solutions.

The perspective on  such renormalization group flows in terms of an effective ``cobordism theory" is convenient if one wants to make explicit connections to other (known) theories. For instance, in the current paper, we exploit it to realize the connection found in~\cite{Beem:2012yn} between AdS$_{5}$ fixed points and the two-dimensional $SU(2)$ Yang-Mill theory on the Riemann surface $\Sigma^{(\rm IR)}_{g}$. In particular, upon explicitly imposing general natural, physically motivated infrared boundary conditions on the ``cobordism theory", which require the entire system to give vacuum AdS$_{5}$ solutions, we show that the three-dimensional ``cobordism theory" indeed reduces to two-dimensional $SU(2)$ Yang-Mills theory on $\Sigma^{(\rm IR)}_{g}$, and in addition we find that the metric on $\Sigma^{(\rm IR)}_{g}$ is in fact required to be constant curvature.

The organization of this paper is as follows. We start in section~\ref{sec:twist} by introducing the relevant $\cN=1$ twist for M5-branes wrapping a genus-$g$ Riemann surface $\Sigma_g$. In section~\ref{sec:7dsugra}, we briefly introduce the Lagrangian and supersymmetry conditions of the maximally supersymmetric $SO(5)$ gauged supergravity, which is our main tool in the following. We continue in section~\ref{sec:eff3d} by first introducing our ansatz in seven dimensions, and then we discuss the effective three-dimensional ``cobordism theory" that crystallizes upon imposing this ansatz. In section~\ref{sec:eff2d}, we impose physically relevant infrared boundary conditions (leading to AdS$_{5}$ vacua) on our three-dimensional effective theory and show that (to leading order in the infrared limit) it reduces to two-dimensional $SU(2)$ Yang-Mills theory. Finally, in section~\ref{sec:Concl} we discuss some consequences of our results, and propose some future directions.

\section{Four-dimensional $\cN=1$ superconformal field theories from M5-branes}
\label{sec:twist}

We now briefly recall how to obtain $\cN=1$ superconformal field theories upon compactifying the six-dimensional $\cN=(2,0)$ theory on a genus-$g$ Riemann surface $\Sigma_g$~\cite{Benini:2009mz,Bah:2011je,Bah:2011vv,Bah:2012dg} (for a more in depth discussion, we refer the reader to~\cite{Bah:2012dg}).\footnote{In the following we shall restrict to compact Riemann surfaces, \ie~without punctures. It would certainly be very interesting to extend our analysis to Riemann surfaces with punctures, and we shall leave that for future investigation.} To preserve $\cN=1$ supersymmetry requires us to implement a particular topological twist. The six-dimensional $\cN=(2,0)$ theory has $OSp(6,2 | 4)$ superconformal invariance, whose bosonic subgroup is given by $SO(6,2) \times USp(4) \sim SO(6,2) \times SO(5)_R$, the latter factor being the R-symmetry group. To implement the the partial topological twist, we have to embed the $SO(2)$ symmetry of the spin connection of the Riemann surface into the R-symmetry group.

Geometrically, the four-dimensional $\cN=1$ field theory is engineered as a stack of M5-branes on $\mathbb{R}^{3,1} \times \Sigma_g$. More precisely, to preserve $\cN=1$ supersymmetry, the M5-branes are wrapping a calibrated K\"ahler two-cycle $\Sigma_g$ inside a Calabi-Yau threefold (see for instance~\cite{Gauntlett:2003di}). Locally (around the zero-section), the Calabi-Yau threefold can be described by the total space of a complex rank-two vector bundle $\cV_{\mathbb{C}}$ over the Riemann surface $\Sigma_{g}$ with structure group $U(2)$, and the fact that it is Calabi-Yau requires
\be\label{eqn:CYcond}
\det \cV_{\mathbb{C}} \ = \ \cK_{\Sigma_g}\,,
\ee
with $\det \cV_{\mathbb{C}} \to \Sigma_{g}$ the determinant bundle of $\cV_{\mathbb{C}}$, and $\cK_{\Sigma_{g}}$ the canonical bundle of $\Sigma_{g}$. Consequently, we may write the Calabi-Yau threefold as the total space of a bundle $\cK_{\Sigma_g}\otimes \tilde\cV$, with $\tilde\cV$ an $SU(2)$ bundle over the Riemann surface $\Sigma_{g}$. 

For instance, one can recover the $\cN=2$ twisted case (class $\cS$) by picking an appropriate $\tilde \cV$ bundle such that the total space simplifies to $\mathbb{C} \times T^*\Sigma_{g}$. Alternatively, if the structure group of $\tilde\cV$ reduces from $SU(2)$ to $U(1)$, the Calabi-Yau threefold is decomposable, and one arrives at the case discussed in~\cite{Bah:2011vv,Bah:2012dg}, where the Riemann surface $\Sigma_g$ is a calibrated K\"ahler two-cycle in a (local) Calabi-Yau threefold given by the total space of the bundle
\be
\begin{tikzcd}
  \mathbb{C}^{2} \arrow[r] 
    & \cL_1 \oplus \cL_2 \arrow[d] \\
&\Sigma_{g} \end{tikzcd}
\ee
where $\cL_{i}$ are complex line bundles of Chern numbers $c_{1}(\cL_{i}) = n_i$ for $i=1,2$. Thus, in terms of the more general setup, this case is given by $\cV_{\mathbb{C}} = \cL_1 \oplus \cL_2$, and one has to impose the Calabi-Yau condition $\cK_{\Sigma_g} \ = \ \cL_1 \otimes \cL_2$, which translates to 
\be
n_1+n_2 \ = \ 2g-2\,.
\ee
Notice that in this case, there is an associated symmetry $U(1)_{1} \times U(1)_{2}$ acting as phase rotations of the respective fibers of the complex line bundles $\cL_{i}$.

\section{Seven-dimensional gauged supergravity}
\label{sec:7dsugra}

We now introduce our main tool, the seven-dimensional maximally gauged supergravity, which was shown to be given by a consistent truncation of eleven-dimensional supergravity on a four-sphere~\cite{Cvetic:1999xp,Nastase:1999cb,Nastase:1999kf}. Thus, any solution of this theory gives rise to a solution in eleven dimensions, and therefore in the current paper we shall restrict to dealing with this effective theory.

The maximally supersymmetric seven-dimensional $SO(5)$ gauged supergravity was introduced in~\cite{Pernici:1984xx}, and is given by the gauging of an $SO(5)_{g}$ subgroup of the $SL(5,\mathbb{R})$ symmetry on the scalar manifold. Thus, it contains a gauged $SO(5)_{g}$ group together with a local composite symmetry $SO(5)_{c}$. The bosonic field content is given by the metric $g_{\mu\nu}$, $SO(5)_{g}$ Yang-Mills gauge fields $\cA_{(1)}^{ij}$, three-forms $S_{(3)}^{i}$ transforming in the fundamental representation of $SO(5)_{g}$, and fourteen scalars $T_{ij}$ parametrizing the coset $SL(5,\mathbb{R})/SO(5)$, with $T_{ij}$ being a symmetric matrix with $\left| \det T \right|  =  1$.\footnote{Notice in this presentation we have performed a particular gauge choice identifying the gauge symmetry $SO(5)_{g}$ with the composite symmetry $SO(5)_{c}$, and in the following we shall not distinguish the two.} The bosonic Lagrangian of the theory is given by\footnote{For the equations of motions, Bianchi identities and more information on the conventions used throughout this paper, we refer to~\cite{Fluder:2017nww}.}
\bea\label{eqn:7DLagrangian}
{\cal L}_7 &=& 
R_{7}\, {*_71} 
- \frac{1}{4} (T^{-1})_{ij}(T^{-1})_{k\ell}\, {*_7 D T_{jk}}\wedge D T_{\ell i}
- (T^{-1})_{ik}\, (T^{-1})_{j\ell}\, *_7 \mathcal{F}_{(2)}^{ij} \wedge \mathcal{F}_{(2)}^{k\ell} \nn\\
&&
- \frac{1}{2} T_{ij}\, {*_7 S_{(3)}^i}\wedge S_{(3)}^j
+ \frac{1}{2m} S_{(3)}^i\wedge DS_{(3)}^i 
- \frac{1}{2m} \epsilon_{i j k \ell m}\, S_{(3)}^i\wedge \mathcal{F}_{(2)}^{j k}\wedge \mathcal{F}_{(2)}^{\ell m} \nn\\
&&
+ \frac{1}{m} \Omega_{(7)} - V \, *_7 1 \,,
\eea
where $*_7$ is the seven-dimensional Hodge star, and the covariant derivatives and field strength are as follows
\bea
DT_{ij} & \ = \ & \diff T_{ij} + g \mathcal{A}_{(1)}^{ik}\, T_{kj} + g \mathcal{A}_{(1)}^{jk}\, T_{ik}\,,\nn\\
D S_{(3)}^i & \ = \  &\diff S_{(3)}^i + g\, \mathcal{A}_{(1)}^{ij}\wedge S_{(3)}^j\,,\nn\\
\mathcal{F}_{(2)}^{ij} & \ = \ & \diff \mathcal{A}_{(1)}^{ij} + g \mathcal{A}_{(1)}^{ik}\wedge \mathcal{A}_{(1)}^{kj}\,.
\eea
Furthermore, the scalar potential is given by
\bea
V \ = \ \frac{1}{2} m^2 \left(2 T_{ij}\, T_{ij} - (T_{ii})^2 \right)\,,
\eea
and $\Omega_{(7)}$ is a Chern-Simons type term, whose explicit form can be found in~\cite{Pernici:1984xx}.

Finally, we write down the supersymmetry conditions for the seven-dimensional $SO(5)$ gauged supergravity
\bea
0 & = & 
 D_\mu \epsilon + \frac{1}{20} m T \gamma_\mu \epsilon - \frac{1}{40} \left( \gamma_\mu {}^{\nu \rho} - 8 \delta_\mu {}^{\nu}\gamma^{\rho} \right) \Gamma_{ij} \epsilon \, \Pi_I{}^{i} \Pi_J{}^{j} F_{\nu \rho}{}^{IJ} \nn\\
 &&+ \frac{m}{10 \sqrt{3}} \left( \gamma_\mu {}^{\nu \rho \sigma} - \frac{9}{2} \delta_\mu {}^{\nu} \gamma^{\rho \sigma} \right) \Gamma^{i} \epsilon \,\left( \Pi^{-1} \right)_i{}^{I} S_{\nu \rho \sigma \, I} \,, \\
0 & = & 
 \frac{1}{2} \gamma^{\mu} \Gamma^{j} \epsilon \, P_{\mu \, ij} +\frac{1}{2} m \left( T_{ij} - \frac{1}{5} T \delta_{ij} \right) \Gamma^{j}\epsilon + \frac{1}{16} \gamma^{\mu \nu} \left( \Gamma_{kl} \Gamma_i - \frac{1}{5} \Gamma_i \Gamma_{kl} \right) \epsilon \, \Pi_I{}^{k}\Pi_J{}^{l} F_{\mu \nu}{}^{IJ} \nn\\
 &&+ \frac{m}{20 \sqrt{3}} \gamma^{\mu \nu \rho} \left( \Gamma_i{}^{j} - 4 \delta_{i}{}^{j} \right) \epsilon \,\left( \Pi^{-1} \right)_j{}^{I} S_{\mu \nu \rho \, I}  \,,
\eea
where $\Pi_{I}{}^{i}$ are composite scalars related to $T_{ij}$ as follows
\bea
T_{ij} \ = \ \left( \Pi^{-1} \right)_i{}^{I} \left( \Pi^{-1} \right)_j{}^{J} \delta_{IJ} \,,\qquad
T \ = \ \delta^{ij} T_{ij}\,,
\eea
and we introduced symmetric and anti-symmetric composite gauge fields $P_{\mu \, ij}$ and $Q_{\mu \, ij}$ respectively via
\be
 Q_{\mu\, [ij]} + P_{\mu \, (ij)} \ = \ \left( \Pi^{-1} \right)_{i}{}^{I}\left( \delta_{I}{}^{J} \partial_\mu + g A_{\mu \, I}{}^{J} \right) \Pi_{J}{}^{k} \delta_{kj}  \,.
\ee
Lastly, capital letter gamma matrices $\Gamma_i$ are elements in ${\rm Cliff}(5,0)$, lower case ones $\gamma_\mu$ are elements in ${\rm Cliff}(6,1)$, and the covariant derivative acts on the Killing spinors as follows
\be
D_{\mu} \epsilon_a \ = \ \partial_\mu  \epsilon_a +\frac{1}{4} \omega_\mu{}^{mn}\gamma_{mn} \epsilon_a + \frac{1}{4} Q_{\mu \, ij} \left( \Gamma^{ij} \right)_a{}^{b} \epsilon_b \,, \quad a =1,\ldots, 4\,,
\ee
with $\omega_\mu{}^{mn}$ the seven-dimensional spin connection. Finally, the mass parameter $m$ is related to the gauge coupling via 
\be
g \ = \ 2m\, .
\ee

\section{Three-dimensional effective ``cobordism theory"}
\label{sec:eff3d}

We now motivate our ansatz for the seven-dimensional gauged supergravity theory, describing M5-branes wrapping a Riemann surface $\Sigma_{g}$ inside a Calabi-Yau threefold, and subsequently discuss the resulting three-dimensional ``cobordism theory" that arises upon imposing this ansatz.

In the general setup of $\cN=1$ superconformal field theories arising from a twisted compactification of the six-dimensional $\cN=(2,0)$ theory on a Riemann surface $\Sigma_{g}$, $\Sigma_{g}$ is a calibrated two-cycle inside a Calabi-Yau threefold. The directions transverse to the M5-branes arise from four directions tangent, and one (flat) direction normal to the Calabi-Yau threefold. The corresponding normal bundle has $U(2)$ structure group, and the condition for supersymmetry is given by requiring (locally) vanishing first Chern class of the total space, which translates to a relation between the Chern classes of the normal bundle and the one of the tangent bundle of $\Sigma_{g}$. Thus, for our (local) supergravity analysis in the seven dimensionoal $SO(5)$ gauged supergravity, we are expected to decompose $SO(5) \to SO(4) \to U(2)$. In order to implement the particular twist, we write $U(2) \sim U(1) \times SU(2)$, and identify the $SO(2)$ spin connection of the Riemann surface with the gauge fields of a particular combination of $U(1)$ factors inside $U(2)$. 
With this in mind, let us introduce our ansatz.

We start with the ansatz for the seven-dimensional metric
\be\label{eqn:7dmetricansatz}
 \diff s^{2}_7 \ = \ e^{-4 \varphi } \diff s_{M_3}^{2} + e^{2 \varphi}\diff s_{\mathbb{R}^{3,1}}^{2} \,,
\ee
where $\diff s^{2}_{M_3}$ is an arbitrary metric on a three-manifold $M_{3}$, the field $\varphi$ depends on coordinates of $M_3$, and
\be
\diff s_{\mathbb{R}^{3,1}}^{2} \ = \ \diff \vec{x}^{2}-\diff t^{2} \,.
\ee
We require that the renormalization group direction resides within $M_{3}$ and in particular in the infrared of the flows, we want $M_{3}$ together with $\mathbb{R}^{3,1}$ to turn into a ``uniformized" metric on the Riemann surface $\Sigma_g$ together with AdS$_{5}$ (see Figures~\ref{Fig:RGUVIR} and~\ref{Fig:RGIRIR} for two explicit examples of flows), 
\be
\mathbb{R}^{3,1} \times M_{3}  \ \overset{\rm IR}{\leadsto} \ {\rm AdS}_{5} \times \Sigma_{g}^{\rm (IR)} \,.
\ee

Furthermore, we are only turning on $U(2) \subset SO(5)$ gauge fields along $M_{3}$. Thus, we embed $U(2) \hookrightarrow SO(4) \hookrightarrow SO(5)$, or more precisely their corresponding Lie algebras.\footnote{The particular embedding does not matter for our purposes, but in order to match to the truncation in~\cite{1999PhLB..457...39L} (and the solutions of~\cite{Bah:2011vv,Bah:2012dg}) one is required to make a particular choice.} Therefore, our ansatz for the $SO(5)$ gauge fields reads
\be
\mathcal{A}_{(1)}^{ab} \ = \ A^{ab} \,, \quad \text{for} \quad a,b \in \{1,\ldots,4\} \,, 
\ee
with the remaining components turned off, and $A^{ab}$ are in fact $U(2)$ gauge fields embedded into $SO(4)$ (\ie~$\mathfrak{u}(2)$-valued one-forms embedded into $\mathfrak{so}(4)$-valued one forms). We emphasize that the gauge fields are only dependent on the (three) coordinates on $M_3$.

Given the reduction of $SO(5)$ to $SO(4)$, we embed the scalars $T_{ij}$ parametrizing the coset $SL(5,\mathbb{R})/SO(5)$ into scalars $\cT$ parametrizing $SL(4,\mathbb{R})/SO(4)$. Thus, we pick the following ansatz for the scalars
\be
 T_{a b} \ = \ e^{\lambda} \mathcal{T}_{ab} \,, \qquad T_{55} \ = \  e^{-4 \lambda} \,, \qquad T_{a 5} \ = \ T_{5a} \ = \   0 \,,
\ee
with $\mathcal{T}_{ab}$ a symmetric unimodular $4\times 4$ matrix, and $\lambda$ as well as $\mathcal{T}_{ab}$ only dependent on the coordinates of $M_{3}$. We now further simplify the ansatz for the scalars, by imposing that $\cT$ and consequently $T$ are diagonal, \ie
\be
T_{ab}  \ = \ {\rm diag} \left( e^{2\lambda_1}, e^{2\lambda_2}, e^{2\lambda_3}, e^{2\lambda_4}, e^{-2(\lambda_1+\lambda_2+\lambda_3+\lambda_4)} \right) \,.
\ee
This allows us to introduce purely $U(2)$ indices for the gauge fields, \ie~$A \ = \ A^{I}T^{I}$, with $T^{I}$ generators of the Lie algebra $\mathfrak{u}(2)$. However we shall mostly use the real $\mathfrak{so}(4)$ notation, \ie~$A= A^{ab} O_{ab}$, with $O_{ab}$ the $\mathfrak{u}(2)$ generators embedded into $\mathfrak{so}(4)$.

Finally, with the above choice of ansatz for the gauge fields and the scalar, we can trivially solve the equation of motion for the three-form $S^{i}$ by setting
\be
S_{(3)}^{i}  \  = \  0\,, \qquad i=1,\ldots, 5 \,.
\ee

Apart from this, we also require the theory to preserve $\cN=1$ supersymmetry in four-dimensions. Thus, we impose the following projection conditions on the Killing spinors~\cite{Maldacena:2000mw}
\be
\gamma_{67} \epsilon_{a} \ = \ \ii \epsilon_{a} \,, \quad
(\Gamma_{12})^{ab} \epsilon_{b} \ = \ \ii \epsilon_{a}\,,\quad 
(\Gamma_{34})^{ab} \epsilon_{b} \ = \ \ii \epsilon_{a}\,, \quad
 \gamma_{5} \epsilon_{a} \ = \ \epsilon_{a} \,,
\ee
with $ a\in \{1, \ldots ,4\} $, which are consistent with the particular $\cN=1$ twist. We shall in the following assume that the Killing spinors $\epsilon_{a}$ surviving the above projection conditions are non-vanishing and solely dependent on the coordinates of $M_{3}$.

Now, let us reduce the seven-dimensional theory encoded in the Lagrangian~\eqref{eqn:7DLagrangian} using the ansatz outlined in the previous section. This three-dimensional effective theory is expected to describe the cobordism between the ultraviolet Riemann surface (with arbitrary metric) to the uniformized infrared Riemann surface, sketched in Figure~\ref{Fig:RGUVIR}. Furthermore, we believe that the generality of the ansatz also allows for this three-dimensional theory to describe flows between the different AdS$_{5}\times \Sigma_{g}^{({\rm IR})_{i}}$ fixed point, as sketched in Figure~\ref{Fig:RGIRIR}.

A careful analysis of the Lagrangian and the corresponding equations of motion and Einstein equation gives the following effective three-dimensional Lagrangian $\cL_{(\rm CB)}$ of the ``cobordism theory" living on the three-manifold $M_{3}$
\bea\label{eqn:eff3dLag}
\cL_{(\rm CB)} (M_{3}) & \ = \ &
R_{3} \, \left(*_{3} 1\right)
-12 \left( *_{3} \diff \varphi \right) \wedge \diff \varphi 
- e^{2 (2 \varphi- \lambda_i - \lambda_j)} \left( *_{3} F^{ij} \right) \wedge F^{ij}\nn\\
&&
\hspace{.4 in}
- e^{2(\lambda_i+\lambda_j)} \left( *_{3} D \Lambda_{ij}  \right)\wedge D \Lambda^{ij} 
- \sum_{i,j=1}^{4}\left(  *_{3} \diff \lambda_i  \right)\wedge \diff \lambda_j\nn\\
&&
\hspace{.4 in}
 - \, \frac{g^{2}}{4}\, V_{3} \, \left(*_{3} 1\right) \,,
\eea
where $R_{3}$ is the three-dimensional Ricci scalar of $M_{3}$, $*_3$ is the three-dimensional Hodge star operator, we introduced the notation
\bea
D \Lambda_{ij} & \ = \ & \delta_{ij}\, e^{-2\lambda_i}\, \diff \lambda_i + \frac{g}{2} \left( e^{-2\lambda_i} - e^{-2\lambda_j}\right) A^{ij} \,, \quad i =1, \ldots 4 \,, 
\eea
for the covariant derivative of the scalars, and the scalar potential is given by
\bea
V_{3} & \ = \ & e^{-4\varphi} \Bigg(
\frac{1}{2}  \sum_{i=1}^{4} e^{4 {\lambda_i}}
+\frac{1}{2}  e^{-4 ({\lambda_1}+{\lambda_2}+{\lambda_3}+{\lambda_4})}\nn\\
&&
\hspace{.6 in}-e^{-2 ({\lambda_1}+{\lambda_2}+{\lambda_3}+{\lambda_4})} \sum_{i=1}^{4}e^{2 {\lambda_i}} 
- \sum_{i > j}^{4} e^{2 ({\lambda_i}+{\lambda_j})}\Bigg)
\,.
\eea
Furthermore the $U(2)$ gauge field (still embedded into $SO(4)$) $A^{ij}$ and its field strength $F^{ij}$ are now fields of a purely three-dimensional theory.

\section{Two-dimensional Yang-Mills at the infrared fixed point}
\label{sec:eff2d}

We shall now take the ``infrared limit" of our effective three-dimensional ``cobordism theory" living on $M_3$. Given the renormalization group flows $\cR_1$ and $\cR_2$ depicted in Figures~\ref{Fig:RGUVIR} and~\ref{Fig:RGIRIR}, we impose that in the infrared limit we get AdS$_{5}\times \Sigma_{g}^{(\rm IR)}$ solutions. Thus, the infrared ``boundary conditions" for $\varphi$ are fixed as follows\footnote{For any real function $p(x)$, we define $p \in o(q)$, provided $p(x)/q(x) \to 0$ for $x\to \infty$.}
\be
\varphi \ = \ g_{(\rm IR)} - \log r + o(1) \,,
\ee
where by $o(1)$ we denote terms that vanish at $r \to \infty$, and $g_{(\rm IR)}$ is a function dependent on the coordinates on $\Sigma_{g}^{(\rm IR)}$. More concretely, the metric on $M_{3}$ is to leading order in the infrared limit given by
\be
( e^{-g_{(\rm IR)}} r )^{4} \, \diff s^{2}_{M_{3}} \ \sim \  \frac{e^{2g_{(\rm IR)}}}{r^{2}} \, \diff r + e^{2 h_0} \, \diff s^{2}(\Sigma_{g}^{(\rm IR)}) \,,
\ee
where $h_0$ is a constant. This simply yields a metric of the form AdS$_{5}\times \Sigma_{g}^{(\rm IR)}$ starting from the seven-dimensional metric ansatz in equation~\eqref{eqn:7dmetricansatz}.

Furthermore, the field strength and composite scalars satisfy the following asymptotic boundary conditions
\be
F^{ij} \ = \ F_{(\rm IR)}{}^{ij}+ o(1) \,, \qquad T^{ij} \ = \ T_{(\rm IR)}{}^{ij} + o (1)\,,
\ee
with $F_{(\rm IR)}^{ij}$ the (non-Abelian) field strength of $A^{ij}_{(\rm IR)}$, a $U(2)$ gauge field on $\Sigma_{g}^{(\rm IR)}$, and $T_{(\rm IR)}$ composite scalars, both of which are only dependent on the coordinates on $\Sigma_{g}^{(\rm IR)}$.

Given these boundary conditions, one can solve the resulting equations of motion and supersymmetry conditions of the seven-dimensional gauged supergravity. We have done so by implementing the extensive system of equations into Mathematica, and by doing so, a somewhat tedious analysis shows that for the scalars, we require\footnote{In the notation of the section~\ref{sec:eff3d}, this translates to $D\Lambda_{ij} = 0 = \diff \lambda_i $.}
\be\label{eqn:TAcommute}
\left[  T_{(\rm IR)} , A_{(\rm IR)}  \right] \ = \ 0 \,, 
\ee
as well as
\be\label{eqn:dT}
\diff T_{(\rm IR)} \ = \ 0 \,.
\ee
Furthermore, the explicit equations imply that the $\Sigma_{g}^{(\rm IR)}$ metric determined by the function $g_{({\rm IR})}$ has constant curvature.\footnote{The fact that $\Sigma_{g}$ ``uniformizes" to a constant curvature metric in the infrared is not that surprising. Indeed, such a statement was explicitly derived (in a less general setting) in~\cite{Anderson:2011cz} and conforms with the field theory intuition~\cite{Gaiotto:2009we}.}

Thus, we conclude that the three-dimensional effective Lagrangian in equation~\eqref{eqn:eff3dLag} reduces (up to some constant factors) to the two-dimensional Lagrangian\footnote{We emphasize again that this statement is actually rather non-trivial, since the equations~\eqref{eqn:TAcommute} and~\eqref{eqn:dT}, as well as the fact that the metric on $\Sigma_{g}^{(\rm IR)}$ is constant curvature, derived from the supergravity equations are non-trivial and crucial in concluding the reduction to two-dimensional Yang-Mills on $\Sigma_{g}^{(\rm IR)}$.}
\be
\cL_{(\rm IR)} ( \Sigma_{g}^{(\rm IR)}) \ \propto \  *_{2} \, F_{(\rm IR)}{}^{I} \wedge F_{(\rm IR)}{}^{I} + o(1) 
\ee
on the Riemann surface $\Sigma_{g}^{(\rm IR)}$. This corresponds to two-dimensional $U(2)$ Yang-Mills theory on $\Sigma_{g}^{(\rm IR)}$. We have neglected the (constant) scalars and metric factors in the theory, as they are explicitly determined from solutions of two-dimensional Yang-Mills theory by the corresponding equations of motion.

Thus, we observed that infared AdS$_{5} \times \Sigma_{g}^{(\rm IR)}$ solutions to our seven-dimensional holographic M5-brane setup reduce to two-dimensional $U(2)$ Yang-Mills theory, a theory with known classical solutions~\cite{Atiyah:1982fa}.\footnote{Indeed, the supergravity constraints fix the remaining fields in terms of the gauge fields, and additionally imposing the Calabi-Yau condition~\eqref{eqn:CYcond}, one arrives at the solutions of~\cite{Bah:2011vv,Bah:2012dg}.} Notice that we have yet to impose the Calabi-Yau condition~\eqref{eqn:CYcond}, but before doing so, we shall discuss gauge-inequivalent solutions of the $U(2)$ Yang-Mills theory on a (compact) Riemann surface $\Sigma_{g}^{(\rm IR)}$ following~\cite{Atiyah:1982fa,Beasley:2005vf,Caporaso:2006kk}.

The critical loci of two-dimensional Yang-Mills theory obviously contain flat connections on $\Sigma_{g}^{(\rm IR)}$. However, there are further (unstable) loci given by solutions to the two-dimensional Yang-Mills equation with non-zero curvature. In order to find such (gauge-inequivalent) solutions, we start by writing the two-dimensional Yang-Mills equation as 
\be\label{eqn:2dYMeqn}
\diff_{A} f \ = \ 0 \,, \quad \text{with} \quad f \ = \  *_{2} F_{A}\,,
\ee
where $\diff_{A}$ is the covariant derivative with respect to the gauge connection $A$ of the $U(2)$-bundle $\cV_{\mathbb{C}}$ over the Riemann surface, and $F_{A}$ is the corresponding curvature two-form. Thus, $f$ is a covariantly constant section of the adjoint bundle associated to $\cV_{\mathbb{C}}$.\footnote{Recall that we denoted by $\cV_{\mathbb{C}}$ the complex rank-two bundle over the Riemann surface with $U(2)$ structure group, whose total space is a local Calab-Yau threefold.} This implies that the $U(2)$ structure group reduces to the centralizer $C_{U(2)}(f) \subset U(2)$ with respect to $f$. Put in more physical terms, the background curvature breaks the gauge group down to $C_{U(2)}(f)$. Hence, any (non-flat) solution to~\eqref{eqn:2dYMeqn} can be described as a flat connection for the gauge group $C_{U(2)}(f)$, twisted by the constant curvature line bundle associated to the $U(1) \subset U(2)$ generated by $f$. However, (gauge-inequivalent) flat connections are in one-to-one correspondence with 
group homomorphisms from the fundamental group $\pi_{1}(\Sigma_{g}^{(\rm IR)})$ into the structure group of the bundle modulo conjugation. Thus (see Theorem 6.7 of~\cite{Atiyah:1982fa}), gauge-inequivalent solutions to~\eqref{eqn:2dYMeqn} are described by conjugacy classes of the (two-dimensional unitary) representations 
\be
\rho : \Gamma_{\mathbb{R}} \ \to \ U(2) \,,
\ee
with $\rho(\pi_{1} (\Sigma_{g}^{(\rm IR)})) \subset SU(2)$, and  $\Gamma_{\mathbb{R}}$ the central extension of $\pi_{1} (\Sigma_{g}^{(\rm IR)})$ by $\mathbb{R}$.%
\footnote{For $\Sigma_{g}^{(\rm IR)}$ a Riemann surface of genus $g\geq 1$, the fundamental group is generated by $2g$ generators $\{a_1, \cdots , a_g, b_1, \ldots , b_g\}$ with the relation $\prod_{i} \left[ a_{i}, b_{i} \right] = 1$, and the commutator is defined as $\left[ a_{i}, b_{i} \right] = a_i b_i a_i^{-1} b_i^{-1}$. The universal central extension $\Gamma$ of $\pi_{1} (\Sigma_{g}^{(\rm IR)})$ by $\mathbb{Z}$ is defined by introducing a further generator $J$ commuting with $a_i$ and $b_i$, $\forall i$ and with the relation $\prod_{i} \left[ a_i, b_i \right] = J$. Then, the (normal) subgroup of this extension, generated by $J$ is isomorphic to $\mathbb{Z}$. Extending the center of $\Gamma$ from $\mathbb{Z}$ to $\mathbb{R}$ gives the group $\Gamma_{\mathbb{R}}$ via the exact sequence
\be
0 \ \to \ \mathbb{R} \ \to \ \Gamma_{\mathbb{R}} \ \to \ \Sigma_{g}^{(\rm IR)} \ \to \ 0 \,.
\ee} 
Given such a homomorphism $\rho$, the field strength $F^{(\rho)}$ associated to the corresponding gauge field $A^{(\rho)}$ is explicitly written as
\be
F^{(\rho)} \ = \ f^{(\rho)} \otimes \omega (\Sigma_{g}^{(\rm IR)})\,,
\ee
with $\omega (\Sigma_{g}^{(\rm IR)})$ the volume form on $\Sigma_{g}^{(\rm IR)}$, and $f^{(\rho)}$ is an element of $\mathfrak{u}(2)$ given by $\diff \rho : \mathbb{R} \to \mathfrak{u}(2)$. Thus, it remains to find all the possible $f^{(\rho)}$.

If $\rho$ is an irreducible representation, $f^{(\rho)}$ is simply given by $\mu \mathbbm{1}_{2} \in \mathfrak{u}(2)$, with $\mu \in \mathbb{R}$. However the Chern class of a principal $U(2)$-bundle is integral, 
\be
\frac{1}{2\pi} \int_{\Sigma_{g}^{(\rm IR)}} \tr \, F^{(\rho)} \ \in \ \mathbb{Z} \,,
\ee
and thus we conclude that 
\be\label{eqn:FrhoU2}
F^{(\rho)} \ = \ \left(\begin{array} {cc} \pi m \ & 0 \\ 0 & \  \pi m \end{array}\right) \  \omega (\Sigma_{g}^{(\rm IR)})\
\ee
for some integer $m \in \mathbb{Z}$.

However, in the case when $\rho$ is (maximally) reducible, $f^{(\rho)}$ is central with respect to a subgroup $U(1) \times U(1) \subset U(2)$. Since $f^{(\rho)}$ is constant, its adjoint action determines a bundle map ${\rm ad} (\cV_{\mathbb{C}}) \to {\rm ad} (\cV_{\mathbb{C}})$ via ${\rm ad}_{f^{(\rho)}}(\, \cdot \, ) =  [ f^{(\rho)}, \, \cdot \ ]$. Thus, we can decompose ${\rm ad} (\cV_{\mathbb{C}})$ into subbundles associated to different eigenvalues of ${\rm ad}_{f^{(\rho)}}$. In particular, this means that the original $U(2)$ bundle is decomposed into a $U(1)_{1} \times U(1)_{2}$-bundle, 
\be
\cV_{\mathbb{C}} \ \to \ \cL_{1} \oplus \cL_{2} \,.
\ee
Now the respective Chern classes for the $\cL_{i}$ are given by the integers $n_{i} \in \mathbb{Z}$, with the total Chern class being their sum $n_1 + n_2$. Thus, we find that 
\be\label{eqn:FrhoU1U1}
F^{(\rho)} \ = \ \left(\begin{array} {cc} 2\pi n_1 \ & 0 \\ 0 & \  2\pi n_2 \end{array}\right)  \ \omega (\Sigma_{g}^{(\rm IR)}) \,.
\ee

However, we have yet to impose the Calabi-Yau condition~\eqref{eqn:CYcond}. This will impose the condition 
\be
c_{1}(\cV_{{\mathbb{C}}}) \ = \ 2 g - 2\,,
\ee 
and thus it will lead to the constraint
\be\label{eqn:CYconstraintChern0}
m \ = \ 2g-2 \,,
\ee
in the case of irreducible representation, and
\be\label{eqn:CYconstraintChern}
n_1 + n_2 \ = \  2 g - 2\,,
\ee 
for reducible representations $\rho$. In particular this conditions further implies that we end up with a two-dimensional $SU(2)$ Yang-Mills theory on the Riemann surface $\Sigma_{g}^{(\rm IR)}$. 

Given this analysis, we conclude that starting from a rather general ansatz, the infrared $\cN=1$ AdS$_{5}$ solutions arising from M5-branes wrapping K\"ahler two-cycles inside a Calabi-Yau threefold are in fact in correspondence with the critical points of $SU(2)$ two-dimensional Yang-Mills theory on $\Sigma_{g}^{(\rm IR)}$. Let us briefly remark on some straightforward consequences of this relation (we shall remark on some more speculative consequences in our conclusion). The following connections between the $\cN=1$ superconformal field theories dual to the AdS$_{5}$ fixed points and two-dimensional $SU(2)$ Yang-Mills have been observed in the reference~\cite{Beem:2012yn}, where the authors study the corresponding superconformal indices and use it to determine the number of relevant and marginal deformations of the fixed points. The latter was already derived in~\cite{Bah:2012dg}.
\begin{itemize}
\item The non-Abelian $U(2)\subset SO(4)$ field strength of the seven-dimensional gauged supergravity reduces to the ones in~\eqref{eqn:FrhoU2} and~\eqref{eqn:FrhoU1U1} (up to some constant normalization and scalar factors) embedded into the Lie algebra $\mathfrak{so}(4)$, and  imposing the constraints~\eqref{eqn:CYconstraintChern0} and~\eqref{eqn:CYconstraintChern}. Notice that in the former case, when $\rho$ is irreducible, only the central $U(1)$ part of $U(2) \sim SU(2) \times U(1)$ is fixed, and we are actually dealing with flat $SU(2)$ connections. This case corresponds to the $\cN=1$ Maldacena-N\'u\~nez solutions, whereas another extremal case given by $n_1=0$ or $n_2=0$ gives the enhanced $\cN=2$ Maldacena-N\'u\~nez solutions~\cite{Maldacena:2000mw}.
\item The dimension of the conformal manifold of the corresponding $\cN=1$ superconformal field theories is given by the sum of the dimension of the complex structure moduli space of the Riemann surface $\Sigma_{g}^{(\rm IR)}$ and the dimension of the moduli space of the critical (stable and unstable) points of the $SU(2)$ Yang-Mills theory on $\Sigma_{g}^{(\rm IR)}$.\footnote{Generically the supergravity moduli give the dimension of a submanifold of the field theory conformal manifold~\cite{Tachikawa:2005tq}, however in this case one can show that they are in fact equal by computing the dimension independently from either side. We thank Y. Wang for pointing this out.} Thus, 
\be
\dim \cM_{(n_1,n_2)} ({\rm CFT}) \ = \ \dim \cM (\Sigma_{g}^{(\rm IR)}) +\dim \cM_{\rho, (n_1, n_2)} (\cV_{\mathbb{C}}) \,,
\ee
where $\cM_{(n_1,n_2)} ({\rm CFT})$ is the conformal manifold of the $\cN=1$ superconformal field theories of~\cite{Bah:2012dg}, labelled by the integers $(n_1,n_2)$, $\cM (\Sigma_{g}^{(\rm IR)})$ is the complex structure moduli space of the Riemann surface $\Sigma_{g}^{(\rm IR)}$, and $\cM_{\rho, (n_1, n_2)} (\cV_{\mathbb{C}})$ is the moduli space of the critical points labelled by the representation $\rho$ as well as the Chern numbers $(n_1, n_2)$.
\item The number of relevant deformations of the $\cN=1$ superconformal field theories dual to the AdS$_{5}$ fixed points (see~\cite{Beem:2012yn}), 
\be
(g-1)+\left| n_{1}-n_{2} \right|
\ee
is precisely reproduced by the Morse index of critical $SU(2)$ connections.
\end{itemize} 

\section{Conclusions and outlook}
\label{sec:Concl}

In this paper we introduced an alternative viewpoint on the holographic setup of M5-branes wrapping nontrivial calibrated cycles inside special holonomy manifolds. We focused on the example of M5-branes wrapping a Riemann surface $\Sigma_{g}$, with a particular twist, leading to $\cN=1$ superconformal field theories in four dimensions. This corresponds to the Riemann surface $\Sigma_{g}$ wrapping a K\"ahler two-cycle inside a Calabi-Yau threefold. We then set up a proper physically motivated ansatz for the effective seven-dimensional $SO(5)$ gauged supergravity describing this particular system with the corresponding twist. This led to a three-dimensional effective theory on the cobordism $M_{3}$ (see Figure~\ref{Fig:RGUVIR} and Figure~\ref{Fig:RGIRIR}), which we believe to describe flows $\cR_1$ between AdS$_{7}$ and AdS$_{5}$ and flows $\cR_{2}$ between different AdS$_{5}$ fixed points. Finally, upon imposing proper infrared boundary conditions one can show that we precisely land on two-dimensional Yang-Mills theory on the infrared Riemann surface, which further is required to be of constant curvature metric.

An obvious next step is to explicitly attempt a construction of the renormalization group flows $\cR_{2}$ (see Figure~\ref{Fig:RGIRIR}) between different AdS$_{5}$ infrared fixed points. We believe that such flows should be described in terms of the effective ``cobordism theory" outlined in section~\ref{sec:eff3d}. In fact it might be fruitful to embed this effective theory into an already known and studied three-dimensional (super)gravity theory, and use possibly known results to conclude the structure of the cobordism $M_3$. Furthermore, the fact that we explicitly observe the Yang-Mills Lagrangian in the infrared is suggestive that this part of the theory remains unchanged along those flows. However, the explicit solutions in~\cite{Bah:2011vv,Bah:2012dg} suggest that the scalars as well as the metric undergo a nontrivial profile when flowing from one fixed point to another.

A straightforward generalization of our treatment in this paper is to explore whether two-dimensional Yang-Mill theory also appears in the infrared when one adds punctures on the Riemann surface (see the references~\cite{Gaiotto:2009gz,Bah:2013wda,Bah:2015fwa} for supergravity duals of the six-dimensional $\cN=(2,0)$ theory reduced on Riemann surfaces with punctures). The classical solutions and moduli spaces of two-dimensional Yang-Mills theory on non-compact Riemann surfaces with punctures are more involved. Thus, employing two-dimensional Yang-Mills on Riemann surfaces with punctures to find a possible classification of AdS$_{5}$ fixed points could be very interesting.

Finally, it is interesting to perform a corresponding analysis for the case of M5-branes wrapping $n$-dimensional manifolds $M_n$ with $n>2$. Preliminary results suggest that the effective theory living on the internal $M_{n}^{(\rm IR)}$ similarly reduces to a Yang-Mills theory with gauge group given by the structure group of the bundle over $M_{n}$, with the total space of the bundle locally describing the special holonomy manifold. In the future, we intend to study in more detail three-manifolds $M_{3}$ inside a $G_2$-manifold and four-manifolds $M_{4}$ inside a $Spin(7)$-manifold, since in those cases the corresponding (normal) bundle is particularly interesting (and nontrivial).

\section*{Acknowledgments}

\noindent 
The author especially thanks Sergei Gukov, for several interesting and important discussions. We further thank Ying-Hsuan Lin and Du Pei for helpful discussions, and Ying-Hsuan Lin and Yifan Wang for comments on the draft. The work of MF is supported by the David and Ellen Lee Postdoctoral Scholarship and the U.S. Department of Energy, Office of Science, Office of High Energy Physics, under Award Number DE-SC0011632.


\bibliographystyle{JHEP}
\bibliography{refs}

\end{document}